\documentclass[%
 reprint,
superscriptaddress,
 amsmath,amssymb,
 aps,
pra,
floatfix,
]{revtex4-1}

\usepackage{ifpdf}
\usepackage{amsmath} 
\usepackage{amssymb}
\usepackage{amsfonts}
\usepackage{braket}
\usepackage{color}
\usepackage[colorlinks=true,linkcolor=blue,citecolor=blue,breaklinks]{hyperref}% add hypertext capabilities
\usepackage{graphicx}% Include figure files
\usepackage{dcolumn}% Align table columns on decimal point
\usepackage{bm}% bold math
\usepackage[normalem]{ulem}

\newcommand{\la}{\langle}
\newcommand{\ra}{\rangle}

\definecolor{Zcolour}{rgb}{0.992, 0.588, 0.22}
\definecolor{purple}{rgb}{0.5,0,0.5}
\definecolor{brown}{rgb}{0.6,0.2,0}
\definecolor{dkgreen}{rgb}{0,0.5,0}

\usepackage{mathtools}

\DeclarePairedDelimiter\floor{\lfloor}{\rfloor}

\begin{document}

%\preprint{APS/123-QED}

\title{Quantum many-body scar states in two-dimensional Rydberg atom arrays}% Force line 
\author{Cheng-Ju Lin}
\affiliation{Perimeter Institute for Theoretical Physics, Waterloo, Ontario N2L 2Y5, Canada}
\author{Vladimir Calvera}
\affiliation{Department of Physics, Stanford University, Stanford, California 94305, USA}
\affiliation{Perimeter Institute for Theoretical Physics, Waterloo, Ontario N2L 2Y5, Canada}
\author{Timothy H. Hsieh}
\affiliation{Perimeter Institute for Theoretical Physics, Waterloo, Ontario N2L 2Y5, Canada}

\date{\today}% It is always \today, today,

\begin{abstract}
We find exponentially many exact quantum many-body scar states in a two-dimensional PXP model --- an effective model for a two-dimensional Rydberg atom array in the nearest-neighbor blockade regime.  Such scar states are remarkably simple valence bond solids despite being at an effectively infinite temperature, and thus strongly violate the eigenstate thermalization hypothesis.
For a particular boundary condition, such eigenstates have integer-valued energies.
Moreover, certain charge-density-wave initial states give rise to strong oscillations in the Rydberg excitation density after a quantum quench and tower-like structures in their overlaps with eigenstates.
\end{abstract}

\maketitle
%\tableofcontents

\section{Introduction}
Recent progress in cold atom experiments has enabled a wealth of fascinating physics, ranging from the realization of entangled states~\cite{Omran570, Greiner2002,deleseleucObservation2019,lienhard2020realization, zhuGeneration2020} to the exploration of new dynamical regimes~\cite{keeslingQuantum2019,Zhang2017_DPT,chiaroGrowth2019,browaeysManybody2020}. 
In particular, a recent experimental observation~\cite{bernienProbing2017} of anomalous dynamics in a Rydberg atom chain has spurred significant interest in what is now dubbed \textit{quantum many-body scar states}, analogous to the single particle scar states~\cite{PhysRevLett.53.1515,PhysRevLett.73.1613} which show nonergodic features despite being highly excited states.
Long time evolution with a generic chaotic Hamiltonian is expected to asymptote to nearly featureless states, as local observables approach their thermal expectation values.  In sharp contrast, systems with quantum many-body scars exhibit atypical dynamics such as strong oscillations in local observables or nonthermal behavior at long times after a quench from some simple product states.  

Underlying this phenomenology is an unusual interplay between non-ergodic and ergodic features.  The eigenstates of a generic chaotic Hamiltonian are expected to have ``volume-law" entanglement entropy scaling and to satisfy the eigenstate thermalization hypothesis (ETH)~\cite{deutschQuantum1991,srednickiChaos1994,rigolThermalization2008}, i.e., the expectation values of local observables will be dictated by the thermal ensemble at temperature set by the eigenstate energy.  In contrast, quantum many-body scar states are exceptional eigenstates at finite-energy density that do not satisfy ETH  and typically have sub-volume law entanglement entropy scaling.~\cite{shiraishiSystematic2017,moriThermalization2017,turnerWeak2018,turnerQuantum2018,moudgalyaExact2018,moudgalyaEntanglement2018,linExact2019,khemaniSignatures2019,schecterManybody2018,iadecolaQuantum2019,hoPeriodic2019,robinsonSignatures2019,jamesNonthermal2019,moudgalyaQuantum,schecterWeak2019,michailidisSlow,choiEmergent2019,paiDynamical2019,bullSystematic2019,Shiraishi_2019,Cort_s_Cubero_2019,suraceLattice,linSlow,khemaniLocal,sugiuraManybody2019,markNew2019,iadecolaQuantum2020,chattopadhyayQuantum,hudomalQuantum,rakovszkyStatistical,moudgalyaThermalization2019,alhambraRevivals,okTopological2019a,shibataOnsager,markUnified,bullQuantum,lee2020exact,magnifico2019real,zhao2020quantum,moudgalya2020large}  
Unlike many-body localized Hamiltonians~\cite{PhysRevLett.111.127201,PhysRevLett.110.260601,PhysRevLett.109.017202,BASKO20061126,abaninColloquium2019,oganesyanLocalization2007,chandranConstructing2015,husePhenomenology2014,aletManybody2018,nandkishoreManybody2015,chenHow2018}, such scar states coexist with other eigenstates that appear to satisfy ETH.  Hence, systems with scars constitute a distinct intermediate regime between complete localization and complete ergodicity.  

An archetypical example is the one dimensional (1D) effective model for the nearest-neighbor blockaded Rydberg atom chain~\cite{bernienProbing2017}.
In this 1D model, Refs.~\cite{turnerWeak2018,turnerQuantum2018} numerically found (nonexact) scar states, proposed to be responsible for the strong oscillatory dynamics observed in the experiment~\cite{bernienProbing2017} after a quench from a charge-density-wave state.
Subsequently, Ref.~\cite{linExact2019} discovered several exact scar eigenstates with manifestly nonthermal properties (e.g. area law scaling of entanglement) and suggested that the numerical scars can be approximated as quasiparticle excitations on the exact scar states. 

Moreover, there has been tremendous progress in the experimental controls of two-dimensional (2D) atom arrays, including implementations of the quantum gates~\cite{PhysRevLett.114.100503,PhysRevA.92.022336} and arranging lattice geometries via optical tweezers~\cite{PhysRevX.4.021034,labuhnTunable2016,Barredo1021,barredoSynthetic2018}.
Motivated by this experimental progress and the experimental proposals on the 2D Rydberg atom platform~\cite{jiTwodimensional2011,celiEmerging,samajdarComplex}, in this Rapid Communication we show that the 2D effective model describing the nearest-neighbor Rydberg-blockaded array also exhibits scar states and anomalous dynamics, with both similarities and important differences from the 1D model.
For the model on a square lattice with periodic boundary conditions, we find exponentially many exact scar eigenstates at a finite energy density which are remarkably simple product states of dimers [valence bond solids (VBS)].
With a particular open boundary condition, we find entire towers of exact low-entangled eigenstates with an integer-valued energies, thus giving rise to strong oscillations with integer frequency upon quenching from a charge-density-wave initial state. 

\section{Setup and model} 
Consider a Rydberg atom system on a lattice where each site can be either in the atomic ground state $|0\ra$ or Rydberg excitation $|1\rangle$, described by the Hamiltonian
\begin{equation}
H_{\text{Ryd}} = \sum_{i=1}^{N}\left( \Omega X_i + \Delta_i n_i \right) + \sum_{i \neq j}  \frac{V_{i,j}}{2}n_i n_j~,
\end{equation}
where $X_i=|1\ra \la 0 | + |0\ra \la 1|$, $n_i \equiv |1 \ra \la 1|$, and $\Delta_i$ is the strength of the detuning laser at site $i$.
$V_{i,j}=V(R_{ij}/a)$ describes the van der Waals interaction between the Rydberg atoms, with the potential $V(x)=C_6/x^6$, where $R_{ij}=|\vec{r}_{i}-\vec{r}_{j}|$ and $a$ is the lattice constant. (Throughout this Rapid Communication, we use units $a=1$ and $\hbar =1$, and set $\Omega=1$ as the energy unit.)

In the regime where the nearest-neighbor interaction $V(1) \gg \Omega$, namely the nearest-neighbor Rydberg blockade regime, we can consider the constrained Hilbert space where no nearest-neighbor excitations $|1\ra$ are allowed.
Projecting to this constrained Hilbert space and truncating the longer-range interactions (and ignoring the detuning laser $\Delta_i=0$), one obtains the effective model,
\begin{equation}\label{eqn:HPXP}
    H = \sum_{i=1}^{N}  X_i \left(\prod_{j: R_{ij}=1} P_j \right)~, 
\end{equation}
where the projector $P_j \equiv |0\ra \la 0 |_j$ ensures the nearest neighbors of the $i$-th atom are in the atomic ground state.  This is called the ``PXP'' model in 1D, and we will continue to use that name in 2D. 

In addition to the lattice symmetries, Eq.~(\ref{eqn:HPXP}) has particle-hole spectrum symmetry: The unitary $\mathcal{C}\equiv \prod_{i}Z_i$, where $Z_{i}\equiv |1\ra\la 1| -|0\ra \la 0|$, transforms $\mathcal{C}H\mathcal{C}^{-1}=-H$. 
Thus every eigenstate $|E\ra$ with energy $E\neq 0$ has a corresponding eigenstate $\mathcal{C}|E\ra$ with energy $-E$.

\begin{figure}
    \includegraphics[width=\columnwidth]{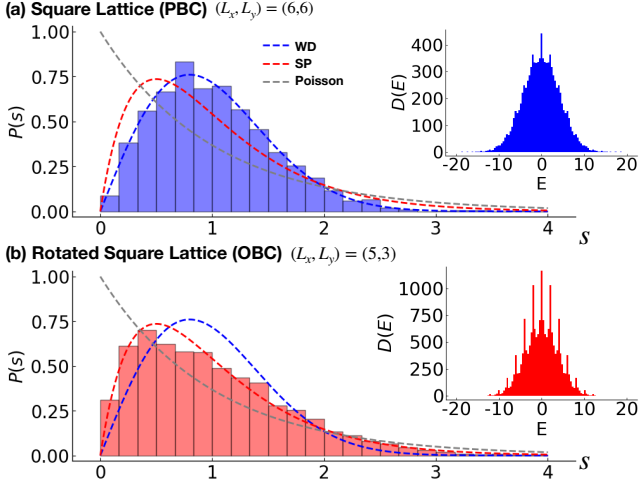}
    \caption{
    (a) Level-spacing statistics of $H$ on a $6 \times 6$ square lattice with a periodic boundary condition (PBC) in the symmetry sector $(K_x,K_y,I_x,I_y,I_{xy})=(0,0,1,1,1)$.
    The statistics is well described by the Wigner-Dyson (WD) statistics $P(s)=\frac{\pi s^2}{2}e^{-\frac{\pi s^2}{4}}$.
    Inset: Density of states.
    (b) Level-spacing statistics of $H$ on a $5 \times 3$ rotated square lattice with an open boundary condition (OBC) in the $(I_x,I_y)=(1,1)$ reflection symmetry sector.
    The statistics is closer to semi-Poisson (SP) statistics $P(s)=4se^{-2s}$.
    Inset: Density of states.
    }
    \label{fig:levelspacing}
\end{figure}

\begin{figure}
    \includegraphics[width=\columnwidth]{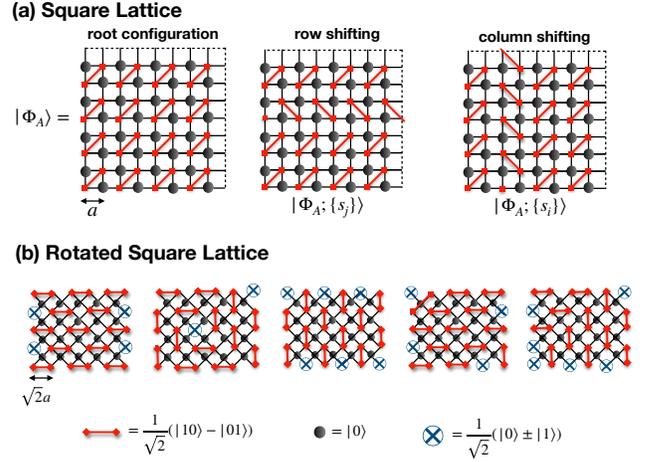}
    \caption{(a) Examples of the exact scar states on a square lattice.
    The dimer configurations can be constructed from the ``root configuration" $|\Phi_{A}\ra$ with the row(column) shifting if the system is periodic in the $x$($y$)-direction.
    (b) Examples of the exact scar states on a rotated square lattice.
    The allowed configurations have at least one pair of dimers on the $A$-sublattice ``freezing" every $B$-sublattice site. 
    The uncovered $A$ sites can be $|s_x=\pm 1 \ra = (|0\ra \pm |1\ra)/\sqrt{2}$, denoted by the $\times$. 
    The eigenenergy is $E=n_{x+}-n_{x-}$, where $n_{x\pm}$ is the number of $|s_x=\pm1 \ra$'s.
    }
    \label{fig:dimerconfig}
\end{figure}

\section{Square lattice}
First, we consider $H$ defined on a square lattice with sites labeled $\{\vec{r}\! =\! (r_x,r_y) | r_{x(y)} = 0 \dots L_{x(y)}\!-\!1 \}$, where dimensions $L_x$ and $L_y$ are even integers.
In Fig.~\ref{fig:levelspacing}, we show the level-spacing statistics of $H$ on a $6 \times 6$ square lattice with a periodic boundary condition (PBC) in the $(K_x,K_y)=(0,0)$ momentum sector and  $(I_x,I_y,I_{xy})=(1,1,1)$ reflection symmetry sector.
[The reflections are defined as $I_x: (r_x, r_y) \rightarrow (L_x\! -\! r_x\!-\!1, r_y)$ and $I_y: (r_x, r_y) \rightarrow (r_x, L_y\! -\! r_y\!-\!1)$.
When $L_x=L_y$ and $I_x=I_y$, one can also assign a quantum number for $I_{xy}: (r_x, r_y) \rightarrow (r_y,r_x)$.]
The statistics is obtained from the unfolded spectrum with eigenindex $n \in [\mathcal{D}/4,\mathcal{D}/2]$ excluding the degenerate states, where $\mathcal{D}$ is the dimension of the Hilbert space in the symmetry sector.
The apparent Wigner-Dyson statistics suggests that the Hamiltonian is not integrable. 
Note that, as shown in the inset, the density of states has a peak at $E=0$; this degeneracy is protected by particle-hole and lattice symmetries~\cite{turnerWeak2018,turnerQuantum2018,schecterManybody2018}.
In addition, the density of states exhibits peaks around some integer-valued energies.

Despite the Wigner-Dyson statistics, we show below that there are simple exact eigenstates in the $E=0$ manifold with area law entanglement.  Define the two sublattices $A(B) = \{\vec{r}| r_x+r_y \in \text{even}(\text{odd})\}$.
We consider a ``root" configuration with $|0\ra$ on the entire $B$ sublattice and dimers $|\chi_{\vec{r},\vec{s}}\ra \equiv \frac{1}{\sqrt{2}}(|1\ra_{\vec{r}}|0\ra_{\vec{s}} - |0\ra_{\vec{r}}|1\ra_{\vec{s}})$ tiling the $A$ sublattice as follows, 
\begin{equation}
    |\Phi_A \ra \equiv \bigotimes_{\vec{a} \in A'}|\chi_{\vec{a},\vec{a}+(1,1)}\ra \bigotimes_{\vec{b} \in B} |0\ra_{\vec{b}}~,
\end{equation}
where $A' =\{(2i,2j) | i(j)=0 \dots L_{x(y)}/2\!-\!1\}$. 
The state $|\Phi_A \ra$ is depicted in Fig.~\ref{fig:dimerconfig}.

It is straightforward to verify $H |\Phi_A \ra=0$. 
First we consider the terms $X_{i}\prod_{j : R_{ij}=1} P_j$ for $i \in B$. 
By design of the dimer configuration, there is at least one excitation $|1\ra$ on a nearest neighbor of $i \in B$, resulting in $X_{i}\prod_{j : R_{ij}=1} P_j|\Phi_A\ra = 0$ for $i \in B$. 
Since the atoms in $B$ are fixed to $|0\ra$, the remaining terms in the Hamiltonian are simply $\sum_{i \in A} X_i$, and 
$(X_{\vec{a}}+X_{\vec{a}+(1,1)})|\chi_{\vec{a},\vec{a}+(1,1)} \ra=0$, and  therefore $H|\Phi_A\ra =0$.  By switching the roles of $A$ and $B$ we can construct the other root configuration
\begin{equation}
    |\Phi_B \ra \equiv \bigotimes_{\vec{b} \in B'}|\chi_{\vec{b},\vec{b}+(1,-1)}\ra \bigotimes_{\vec{a} \in A} |0\ra_{\vec{a}}~,
\end{equation}
where $B' = \{(2i,2j+1) | i(j)=0 \dots L_{x(y)}/2\!-\!1\ \}$.
$|\Phi_{A(B)}\ra$
are zero energy eigenstates of $H$ on a square lattice regardless of the boundary condition.

Based on these two root configurations, we can construct exponentially many eigenstates given periodic or mixed boundary conditions. 
Consider periodic boundary conditions along the $x$ direction [identifying $(L_x+r_x,r_y)$ with $(r_x,r_y)$].
We can independently ``shift" each row of dimers along the $x$ direction, as depicted in Fig.~\ref{fig:dimerconfig}.  Define the index 
$\{s_j= \pm 1 \}$ denoting whether to shift ($-1$) or not ($1$) for each dimer row $j=0 \dots L_y/2-1$.
The corresponding eigenstates are
\begin{equation}
    |\Phi_{A} ;\{ s_j\} \rangle = \bigotimes_{\vec{a} \in A'}|\chi_{\vec{a},\vec{a}+(s_j,1)}\ra \bigotimes_{\vec{b} \in B} |0\ra_{\vec{b}}~, 
\end{equation}
Similarly, we can construct row-shifted configurations starting from the root configuration $|\Phi_{B}\ra$,
\begin{equation}
    |\Phi_{B} ;\{ s_j\} \rangle = \bigotimes_{\vec{b} \in B'}|\chi_{\vec{b},\vec{b}+(s_j,-1)}\ra \bigotimes_{\vec{a} \in A} |0\ra_{\vec{a}}~. 
\end{equation}
Given periodic boundary conditions in the $y$ direction, one can analogously define ``column-shifting" as depicted in Fig.~\ref{fig:dimerconfig} and specify states $|\Phi_{A} ;\{ s_i\} \rangle, |\Phi_{B} ;\{ s_i\} \rangle$
labeled by the ``column-shifting index" $\{s_i= \pm 1\}$. 

While these states are all eigenstates with energy $E=0$, they are not all mutually orthogonal.
However, we show in the Supplemental Material~\cite{zotero-2074} that the set of states $\{ |\Phi_{A} ;\{ s_{j}\} \rangle , |\Phi_{B} ;\{ s_{j}\} \rangle\}$ is linearly independent (and in fact orthogonal in the thermodynamic limit).  
The same applies to the column-shifted states $\{ |\Phi_{A} ;\{ s_{i}\} \rangle , |\Phi_{B} ;\{ s_{i}\} \rangle\}$.
Therefore, there are at least $2^{\max (L_x,L_y )/2+1}$ linearly independent exact scar states. 
(We note that they do not exhaust the entire $E=0$ manifold.)

Note that these exponentially many states are all at effectively infinite temperature, yet they exhibit VBS order breaking translation invariance and have area-law entanglement ($n \ln 2$ where $n$ is the number of dimers intersecting the bipartition).

\section{Rotated square lattice} 
We next consider $H$ on a square lattice with boundaries cut at $45^{\circ}$ with respect to the lattice axes.  We refer to this geometry as the rotated square lattice [see Fig.~\ref{fig:dimerconfig}(b)]. 
The new coordinates of the sublattices are $A  \equiv \{ \sqrt{2} (i, j) | i(j)=0 \dots L_{x(y)}-1 \}$ and $B  \equiv \{ \sqrt{2} (i, j) + \frac{1}{\sqrt{2}}(1,1) | i(j)=0 \dots L_{x(y)}-2 \}$.
Therefore there are $N= L_x\! \times\! L_y\! +\! (L_x \!-\!1)\! \times\! (L_y\!-\!1)$ total lattice sites.

Interestingly, $H$ on such a lattice geometry has degenerate eigenstates at \textit{integer energies}. For example, in the $(L_x,L_y)=(5,3)$ system, there are degenerate eigenstates at $E= \pm5, \pm3, \pm1$ and $0$.
Note that the degeneracies at $E \neq 0$ are not protected by the symmetries.
The level-spacing statistics of the \textit{nondegenerate} eigenstates with eigenindex $n \in [\mathcal{D}/4, \mathcal{D}/2]$ for this geometry are also shown in Fig.~\ref{fig:levelspacing}; the statistics is described reasonably well by a semi-Poisson distribution, although this could be an artifact of finite size effects as in the 1D model \cite{turnerWeak2018}. 

Specifically, in the inset of Fig.~\ref{fig:levelspacing}, the spectrum exhibits even stronger clustering around the integer-valued energies, which may account for the semi-Poisson statistics.
It is possible that these clustered satellite states are only subextensive and therefore the statistics will approach Wigner-Dyson for larger system sizes.
The nature and density of such satellite states is an intriguing question requiring further studies.    

As in the regular square lattice, we can construct exact eigenstates on the rotated square lattice with integer-valued energies.
We again seek configurations which ``freeze" $B$ sites to $|0 \ra$ by having at least one neighboring dimer.
Yet there are several differences between this rotated geometry and the previous case.  In this case, the dimers need not entirely cover $A$, and residual $A$ sites can be fixed to be $|s_x = \pm 1 \ra = \frac{1}{\sqrt{2}}(|0\ra \pm |1\ra)$.  Moreover, the dimer orientation has additional flexibility; for example, $(1,0),(0,1),(1,1)$ directions are all possible.
Formally, given a dimer covering $V(D)$ where $D$ is the set of lattice sites covered by the dimers, define the states
\begin{equation}
    |V(D),\{s_x\} \ra = |V(D)\rangle \bigotimes_{\vec{a} \in \bar{D}}|s_x \ra_{\vec{a}} \bigotimes_{\vec{b} \in B}|0\ra_{\vec{b}}~,
\end{equation}
where $\bar{D}$ is the complement of $D$ in $A$.  Such an eigenstate has energy $E = (n_{x+}-n_{x-})$, where $n_{x\pm}$ is the number of $|s_x= \pm 1 \ra$'s.
We depict some examples in Fig.~\ref{fig:dimerconfig}(b).
We again expect that these states probably do not exhaust the degeneracies observed numerically, yet we can show that there are at least $O(2^{\max (L_x,L_y)})$ such eigenstates~\cite{zotero-2074}.

\section{Dynamical signatures} 
\begin{figure}
    \includegraphics[width=\columnwidth]{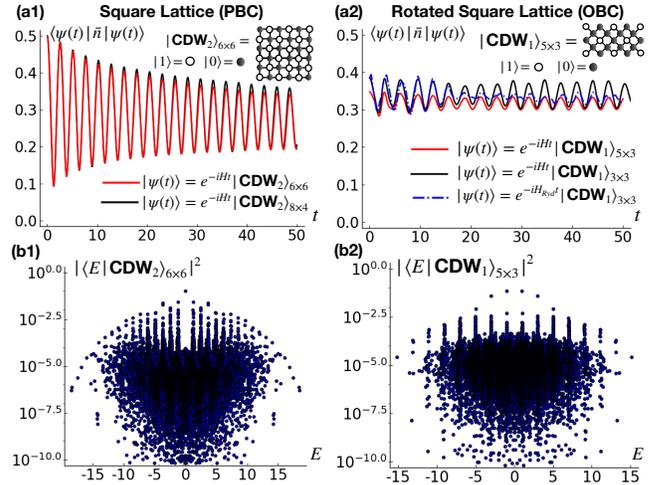}
    \caption{Left panel: On the square lattice (PBC), (a1) quench dynamics of the Rydberg excitation density $\bar{n} \equiv \sum_i n_i / N$ with the charge-density-wave initial states $|\text{CDW}_2\rangle$ under the PXP Hamiltonian $H$ with sizes $6 \times 6$ and $8 \times 4$.
    (b1) The weight distribution of the charge-density-wave state $|\text{CDW}_2\rangle$ on the eigenstates of $H$.
    Right panel: (a2) - (b2) Similar to the figures on the left panel but on the rotated square lattice (OBC) and with initial state $|\text{CDW}_1\rangle$ in sizes $5\! \times \! 3$ ($N\!=\!23$) and $3\! \times\! 3$ ($N\!=\!13$).
    Additionally in (a2), we also show the evolution under $H_{\text{Ryd}}$ with $C_6=4$.}
    \label{fig:dynamics}
\end{figure}

Given these exact scar states on the rotated square lattice with integer-valued energies, we ask if any simple product state can have a high overlap with these states and potentially lead to oscillations in quench dynamics.
Indeed, we consider the charge-density-wave (CDW) state $|\text{CDW}_1 \rangle \equiv \bigotimes_{\vec{r} \in A} |\sigma \rangle _{\vec{r}} \bigotimes_{\vec{b} \in B}|0\rangle_B$, where $|\sigma\! =\! 1 \rangle _{\vec{r}}$ when $\vec{r}\!=\!\sqrt{2}(n_x,n_y)$ if $n_x\!+\! n_y$ is even or $|\sigma \!=\! 0 \rangle _{\vec{r}}$ if $n_x\!+\! n_y$ is odd, as depicted in Fig.~\ref{fig:dynamics}(a2).
Such a CDW state has an overlap with states in the first and third categories depicted in Fig.~\ref{fig:dimerconfig}(b), for example. 

In Fig.~\ref{fig:dynamics}(a2), we see oscillatory dynamics with angular frequency $\omega \approx 2$. 
Since $|\text{CDW}_1\rangle$ has an overlap with the exact scar states we constructed earlier, the oscillatory dynamics is partially accounted for by the exact scar states.
Figure~\ref{fig:dynamics}(b2) shows the weight distribution of $|\text{CDW}_1\rangle$ over the eigenstates. 
It shows a similar tower structure as in the 1D PXP model, and also has relatively higher overlap with the integer-valued degenerate eigenspaces.

However, we stress that the overlaps between $|\text{CDW}_1 \rangle$ and the scar manifold $|V(D),\{s_x\} \ra$ re likely to approach zero in the thermodynamic limit $L_x, L_y \rightarrow \infty$. 
For the first and third types of scar states in Fig.~\ref{fig:dimerconfig}(b), each dimer (and each $|s_x\ra$) contributes $1/\sqrt{2}$ to the overlap; since there are $O(\alpha L_x L_y)$ of them, the overlap with each state is $O(2^{- \gamma L_x L_y})$ with some constants $\alpha$ and $\gamma$.  
However, the number of these exact scar states is likely to be only exponential in the linear dimension, resulting in a zero overlap in the thermodynamic limit.
Nevertheless, for finite sizes, the overlap with the scar manifold $|V(D),\{s_x\} \ra$ will be nonzero.

In Fig.~\ref{fig:dynamics}(a2), we also considered the evolution of $|\text{CDW}_1 \rangle$ under the full Rydberg Hamiltonian $H_{\text{Ryd}}$ with $C_6 =4 $. 
We see that in system size $3 \times 3$, the exact and effective dynamics match up to $t\approx 18$. 
Note that there will be a competition between the blockade-constraint and the longer-range interaction when using $H_{\text{Ryd}}$ to realize $H$.
$C_6 = 4$ in this case seems to be ideal for comparing with $H$, and different $C_6$ will result in a different overall Rydberg excitation density while still having a similar oscillation frequency.

On the other hand, quenching from $|\text{CDW}_1 \ra$ does not show strong oscillatory dynamics on both PBC and regular OBC square lattices.
However, a different CDW state $|\text{CDW}_2\ra  = \bigotimes _{\vec{a} \in A} |1 \rangle _{\vec{a}} \bigotimes _{\vec{b} \in B} |0 \rangle _{\vec{b}}$  does show strong oscillatory dynamics.  In Fig.~\ref{fig:dynamics}(a1), we show the Rydberg excitation density as a function of time $\la \psi(t)|\bar{n}|\psi(t)\ra$, where $\bar{n}\equiv \sum_i n_i /N$ with sizes $(L_x,L_y)=(8,4)$ and $(6,6)$. 
It exhibits strong oscillations with angular frequency $\omega \approx 2.5$. 
Examining the weight distribution of $|\text{CDW}_2 \rangle$ on the eigenstates of $H$ as shown in Fig.~\ref{fig:dynamics}(b1), we see a set of towers almost equally spaced in energy, similar to that of the 1D PXP model.
Note that the oscillation frequency is given by roughly twice the energy spacing between the towers, because the towers are in different symmetry sectors alternating between $(I_x,I_y,I_{xy})=(1,1,1)$ and $(I_x,I_y,I_{xy})=(-1,-1,1)$ consecutively.  The state $|\text{CDW}_2 \rangle$ does not have an overlap with the exact scar states $|\Phi_{A} ;\{ s_j\} \rangle$, etc., however, and thus understanding such oscillatory dynamics requires further studies.

\section{Entanglement entropy and testing ETH}
\begin{figure}
    \includegraphics[width=\columnwidth]{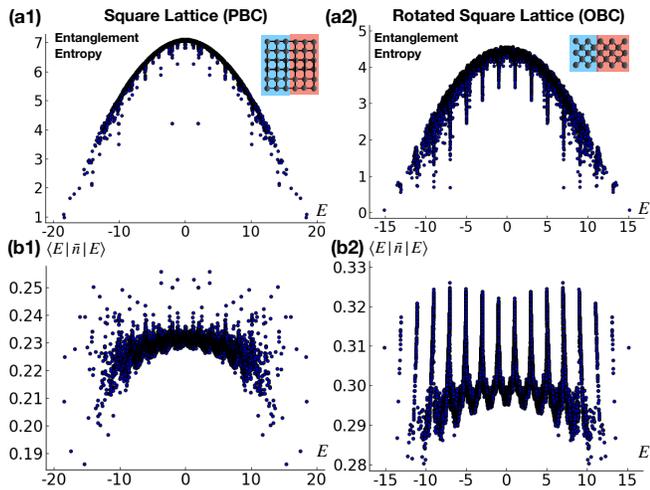}
    \caption{
    Left panel: For the nondegenerate eigenstates of $H$ on a $6 \times 6$ square lattice (PBC), (a1) the entanglement entropy with the bipartition shown in the figure and (b1) the Rydberg excitation density $\bar{n}=\sum_i n_i/N$, in the symmetry sectors $(K_x,K_y,I_x,I_y,I_{xy})=(0,0,1,1,1)$  and $(\pi,\pi,-1,-1,1)$.
    Right panel: (a2)-(b2) Similar to the left panel but on a $5 \times 3$ rotated square lattice (OBC) in all the symmetry sectors.
    }
    \label{fig:ETHEE}
\end{figure}

The anomalous tower structure in the overlaps between the eigenstates and the CDW states warrants a further examination of the properties of the eigenstates.
In Fig.~\ref{fig:ETHEE}, we study the entanglement entropy and a test of ETH in the \textit{nondegenerate} eigenstates in the PBC square and OBC rotated square lattice.

We first discuss the eigenstates on a $6 \times 6$ PBC square lattice in the symmetry sectors $(K_x,K_y,I_x,I_y,I_{xy})=(0,0,1,1,1)$ and $(\pi,\pi,-1,-1,1)$.
In Fig.~\ref{fig:ETHEE}(a1), we see that the entanglement entropy of the majority of the eigenstates (with the bipartition specified in the figure) seems to depend smoothly on the energy, with some apparent outliers showing signs of ETH violation. 
The same holds for the expectation value of the Rydberg excitation density $\bar{n}=\sum_i n_i /N$ of the eigenstates [Fig.~\ref{fig:ETHEE}(b1)].
In fact, we observe that the eigenstates at $E=\pm 2$ are clearly outliers, and the wavefunction display a more intricate translation symmetry breaking pattern than our dimer constructions.
We also note that on an OBC square lattice, all the nondegenerate eigenstates in the middle of the spectrum do not show clear signatures of ETH violation~\cite{zotero-2074}.

The (nondegenerate) eigenstates on a rotated square lattice show even more striking behavior. 
In Fig.~\ref{fig:ETHEE}(a2), towers of the eigenstates near the integer energies show anomalously low entanglement entropy; 
Fig.~\ref{fig:ETHEE}(b2) clearly shows that these states are also outliers in the ETH test with the Rydberg excitation observable. 
These energies correspond to the peaks in the density of states in Fig.~\ref{fig:levelspacing}(b).

\section{Discussions}
We showed several striking similarities and differences between the 2D PXP model and the 1D PXP model.
In the 2D PXP model, there are exact scar states which are ideal VBS states, even simpler than those of the 1D model~\cite{linExact2019}.
Such constructions can be generalized to higher dimensional PXP model~\cite{zotero-2074}.  In addition, the CDW quench dynamics also show strong oscillations and revival signatures, as in the 1d model, and this phenomena can be observed in near-term Rydberg atom experiments with 2D arrays.

However, the 2D model exhibits a sharp dependence on boundary conditions not seen in the 1D counterpart.  In particular, the rotated square lattice geometry contains exact scar eigenstates at integer-valued energies, resulting from the fact that there can be extra $A$-sublattice sites (``monomers") free from the dimer covering to freeze the $B$-sublattice.
In contrast, all the $A$-sublattice sites in the regular square lattice have to be covered by dimers to freeze the $B$-sublattice.
And this energy spacing manifests in the frequency of the anomalous oscillations in quench dynamics, corresponding to the Rabi oscillation of the ``monomers".

Moreover, unlike in the 1d model, the degree of ETH violation for states other than the exact scars seems to greatly depend on the boundary conditions and lattice geometry.
While some eigenstates on the OBC rotated square lattice show strong signs of ETH violation, some eigenstates on the PBC square lattice show weaker signs of ETH violation and the eigenstates on the OBC square lattice show almost no signs of violation~\cite{zotero-2074}.

\textit{Note added: } Recently, we became aware of a paper (Ref.~\cite{michailidisStabilizing2020}) on deforming the 2D PXP model to stabilize oscillatory dynamics.

\begin{acknowledgments}
We thank A.~M.~Alhambra, A.~Chandran, Y.-C.~He, W.~W.~Ho,  D.~K.~Mark, H.~Ma, O.~I.~Motrunich, and A.~Szasz for valuable discussions, and especially A.~M.~Alhambra, A.~Chandran, W.~W.~Ho and O.~I.~Motrunich for invaluable feedback on our manuscript.
C.-J.~L.\ and T.~H.~H.\ acknowledge support from Perimeter Institute for Theoretical Physics and Compute Canada (\href{https://www.computecanada.ca}{www.computecanada.ca}).
V.~C.\ was supported by the Visiting Graduate Fellowship program at Perimeter Institute where part of this work was carried out.
This research was supported in part by Perimeter Institute for Theoretical Physics. 
Research at Perimeter Institute is supported in part by the Government of Canada through the Department of Innovation, Science and Economic Development Canada and by the Province of Ontario through the Ministry of Colleges and Universities.
\end{acknowledgments}

%\bibliography{2DPXP}% Produces the bibliography via BibTeX.

%%%%%%%%%% Merge with supplemental materials %%%%%%%%%%
\clearpage
\onecolumngrid
\begin{center}
\textbf{\large Supplemental Material: Quantum Many-Body Scar States in Two-Dimensional Rydberg Atom Arrays}
\end{center}
%%%%%%%%%% Merge with supplemental materials %%%%%%%%%%
%%%%%%%%%% Prefix a "S" to all equations, figures, tables and reset the counter %%%%%%%%%%
\setcounter{equation}{0}
\setcounter{figure}{0}
\setcounter{table}{0}
\setcounter{page}{1}
\setcounter{section}{0}
\makeatletter
\renewcommand{\theequation}{S\arabic{equation}}
\renewcommand{\thesection}{S\arabic{section}}
\renewcommand{\thefigure}{S\arabic{figure}}
\renewcommand{\bibnumfmt}[1]{[S#1]}
\renewcommand{\citenumfont}[1]{S#1}
%%%%%%%%%% Prefix a "S" to all equations, figures, tables and reset the counter %%%%%%%%%%

\section{Proof of linear independence of the exact scar states}\label{app:Overlap}
In this section, we show that the states $\{ |\Phi_A; \{s_{j} \} \ra,  |\Phi_B; \{s_{j} \} \ra\}$ are linearly independent.
Recall that 
\begin{equation}
    |\Phi_{A} ;\{ s_{j}\} \rangle = \bigotimes_{\vec{a} \in A^{\prime}}|\chi_{\vec{a},\vec{a}+(s_{j},1)}\ra \bigotimes_{\vec{b} \in B} |0\ra_{\vec{b}}~, 
\end{equation}
and
\begin{equation}
    |\Phi_{B} ;\{ s_{j}\} \rangle = \bigotimes_{\vec{b} \in B^{\prime}}|\chi_{\vec{b},\vec{b}+(s_{j},-1)}\ra \bigotimes_{\vec{a} \in A} |0\ra_{\vec{a}}~, 
\end{equation}
where $\{s_{j} = \pm1\}$ is the collection of the ``shifting index", $A^{\prime} =\{(2i,2j) | i(j)=0 \dots L_{x(y)}/2\!-\!1\}$ and $B^{\prime} = \{(2i,2j+1) | i(j)=0 \dots L_{x(y)}/2\!-\!1\ \}$.

Now to calculate the overlap of $|\Phi_A; \{s_{j} \}\rangle$ and $|\Phi_A; \{s_{j}^{\prime} \}\rangle$, consider the dimer row $j$, if $s_{j}=s_{j}^{\prime}$, the contribution of this row to the overlap will just be $1$; if $s_{j}\neq s_{j}^{\prime}$, the contribution will be $2\cdot 2^{-L_x/2}$.
The matrix elements of the $2^{L_y/2} \times 2^{L_y/2}$ overlap matrix is therefore
\begin{align}
 G(\{s'_{j} \},\{s_{j}\})\equiv\langle\Phi_A; \{s'_{j}\}|\Phi_A; &\{s_{j}\} \rangle = \prod_{n_y=0}^{L_y/2-1}[2 \cdot 2^{-L_x/2} ]^{\frac{|s_{j}^{\prime}-s_{j}|}{2}}~,
\end{align}
which is equivalent to
\begin{align}
 G &=\bigotimes_{j=0}^{L_y/2-1}( I_{2\times2} + 2\cdot 2^{-L_x/2}\sigma^x)~,
\end{align}
where $I_{2\times2}$ is a $2 \times 2$ identity matrix.
We therefore obtain $\det (G)=[1-4\cdot 2^{-L_x}]^{L_y/2} \neq 0$, when $L_x > 2$.
In other words, $|\Phi_{A} ;\{ s_{j}\} \rangle$'s are linearly independent.
In the thermodynamics limit $L_x \rightarrow \infty$, $|\Phi_{A} ;\{ s_{j}\} \rangle$'s are in fact orthogonal to each other.
Moreover, since ${|\Phi_A,\{s_{j}\}\rangle}$ and ${|\Phi_B,\{s_{j}^{\prime}\}\rangle}$ are orthogonal, we have at least $2 \cdot 2^{L_y/2}$ exact $E=0$ scar states if we have periodic boundary conditions in the $x$ directions.
The same argument also applies if we exchange $x \leftrightarrow y$.

If we have periodic boundary conditions in both directions, from the previous analysis, we see that there are at least  $2^{\text{max}(L_x,L_y)/2 + 1}$ linearly independent eigenstates. 
Notice that in this case, two extra root configurations can be considered:
\begin{equation}
    |\Phi_A^{\prime} \ra \equiv \bigotimes_{\vec{a} \in A^{\prime \prime}}|\chi_{\vec{a},\vec{a}+(1,1)}\ra \bigotimes_{\vec{b} \in B} |0\ra_{\vec{b}}~,
\end{equation}
and 
\begin{equation}
    |\Phi_B^{\prime} \ra \equiv \bigotimes_{\vec{b} \in B^{\prime \prime}}|\chi_{\vec{b},\vec{b}+(1,-1)}\ra \bigotimes_{\vec{a} \in A} |0\ra_{\vec{a}}~,
\end{equation}
where $A^{\prime \prime} =\{(2i,2j)+(1,1) | i(j)=0 \dots L_{x(y)}/2\!-\!1\}$ and $B^{\prime \prime} = \{(2i+1,2j) | i(j)=0 \dots L_{x(y)}/2\!-\!1\ \}$.
Starting from these configurations, we can also perform column and row shifting, obtaining many more states.
However, notice that when we shift all columns(rows) starting from these new root states, we arrive at the same states as shifting all rows(columns) starting from the original root states.
Therefore, there are $2 \cdot(2^{L_x/2}+2^{L_y/2}-2)$ states defined on the $A$ sublattice.
In the thermodynamic limit $L_x,L_y \rightarrow \infty$, these states are all orthogonal to each other.
We checked that for the $L \times L$ square lattice that these $4 \cdot (2^{L/2}-1)$ states are actually linearly independent for small system sizes $L=4,6,8,10$.

\section{Results on the square lattice with open boundary condition}
\begin{figure}
    \includegraphics[width=0.5\columnwidth]{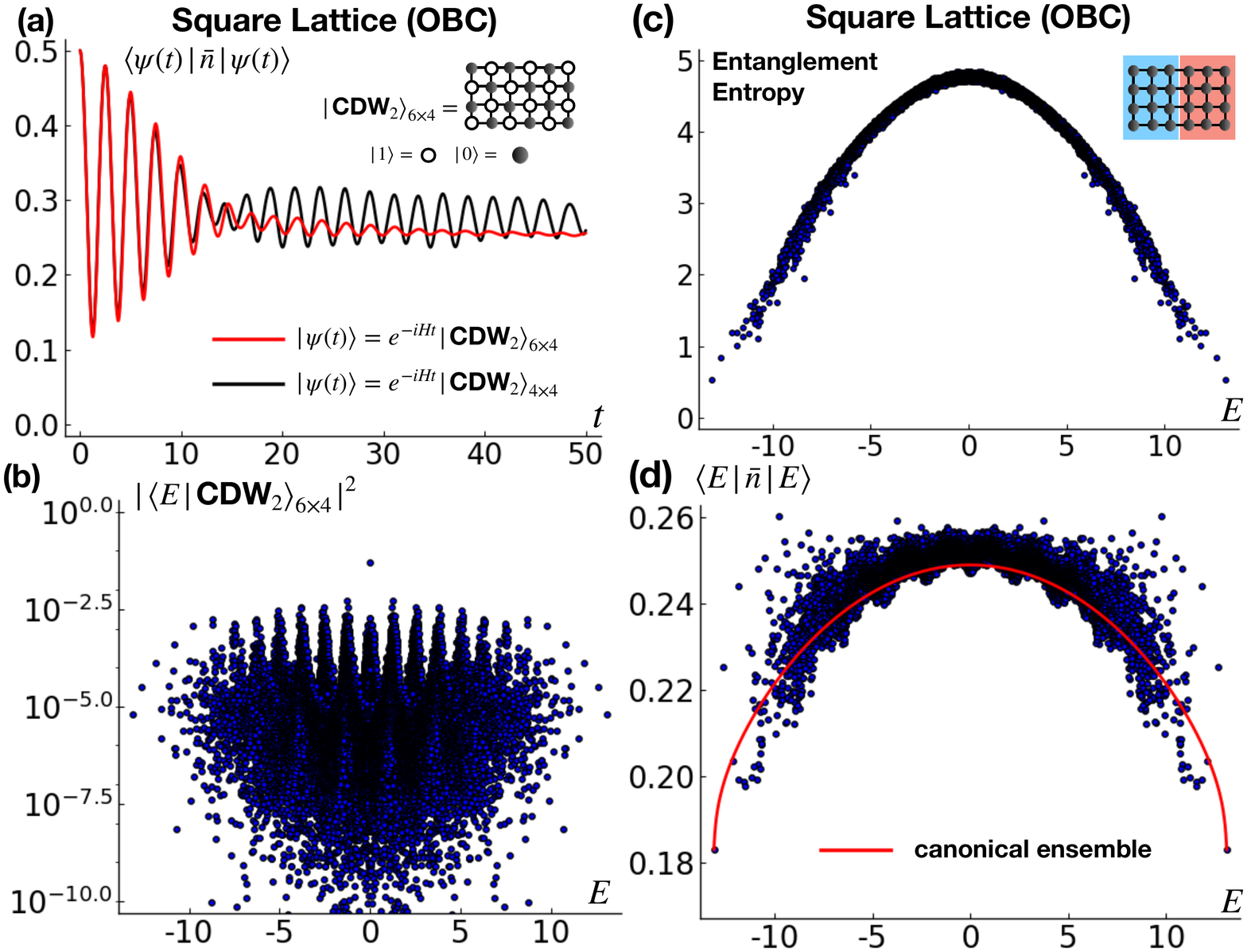}
    \caption{(a)The quench quench dynamics of the Rydber excitation $\bar{n}\equiv \sum_i n_i /N$ starting with $|\text{CDW}_2\ra$ state on the OBC square lattices.
    (b) The overlap distribution of $|\text{CDW}_2\ra$ on the eigenstates of $H$ on the $6 \times 4$ OBC square lattice.
    (c) The bipartite entanglement entropy of the (nondegenerate) eigenstates of $H$ on the $6 \times 4$ OBC square lattice in all the symmetry sectors.
    (d) Rydberg excitation density $\bar{n}$ of the (nondegenerate) eigenstates of $H$ on the $6 \times 4$ OBC square lattice in all the symmetry sectors.
    }
    \label{fig:OBC}
\end{figure}

Here we discuss the results on the square lattice with OBC.
In Fig.~\ref{fig:OBC}(a), we studied the quench dynamics starting from the charge-density-wave state $|\text{CDW}_2 \rangle  = \bigotimes _{\vec{a} \in A} |1 \rangle _{\vec{a}} \bigotimes _{\vec{b} \in B} |0 \rangle _{\vec{b}}$ on a OBC square lattice of size $L_x \times L_y$ with open boundary condition and evolving under $H$.
We show the Rydberg excitation density as a function of time $\la \psi(t)|\bar{n}|\psi(t)\ra$, where $\bar{n}\equiv \sum_i n_i /N$ with sizes $(L_x,L_y)=(4,4)$ and $(6,4)$.
It exhibits strong oscillations with angular frequency $\omega \approx 2.5$. 
Examining the weight distribution of $|Z_2 \rangle$ on the eigenstates of $H$ as shown in Fig.~\ref{fig:OBC}(b), we see a set of towers almost equally spaced in energy, similar to that of the 1d PXP model.
Note that the oscillation frequency is given by roughly twice the energy spacing between the towers, because the towers are in different symmetry sectors alternating between $(I_x,I_y)=(1,1)$ and $(I_x,I_y)=(-1,-1)$ consecutively.

In Fig.~\ref{fig:OBC}(c), we see that the entanglement entropy of the
eigenstates (with the bipartition specified in the figure)
seems to depend smoothly on the energy. 
There are some outliers away from the middle of the spectrum, but this
could be due to finite size effects. 
The same holds for the expectation value of the Rydberg excitation density $\bar{n}=\sum_i n_i/N$ of the eigenstates (Fig.~\ref{fig:OBC}(d)). 
Despite the anomalous structure in the overlaps, the entanglement entropy and the Rydberg excitation density show no signs of ETH violation.

This behavior is certainly unlike the 1d model, the 2d PBC square lattice  or the 2d rotated square lattice model.  
It is possible that a superposition of states in an energy window containing the high overlap states is required to construct a low entanglement state, which is a necessary consequence of approximate revivals in quench dynamics~\cite{alhambraRevivals}.

\section{Exponentially many exact scar states on a rotated square lattice}
Here we show there are also at least exponentially many exact scar states on a rotated square lattice.
Without loss of generality, consider the rotated square lattice with $L_x \geq L_y$, with lattice sites $\Lambda =A \cup B$, where $A  \equiv \{ \sqrt{2} (i, j) | i(j)=0 \dots L_{x(y)-1} \}$ and $=B  \equiv \{ \sqrt{2} (i, j) + \frac{1}{\sqrt{2}}(1,1) | i(j)=0 \dots L_{x(y)-2} \}$.
For a specific dimer covering and the corresponding exact scar states
\begin{equation}
    |\Gamma; \{s_x\} \ra \equiv \bigotimes_{\vec{q} \in Q}|\chi_{\vec{q},\vec{q}+\sqrt{2}(1,0)}\ra \bigotimes_{\vec{b} \in \Lambda_B^{\prime}}|0\ra_{\vec{b}} \bigotimes_{\vec{a} \in \bar{D}} |s_x \ra _{\vec{a}}~,
\end{equation}
where $Q \equiv \{\sqrt{2}(n_x, n_y)| n_x + n_y \in \text{even}; n_x = 0 \dots L_x\! -\!1; n_y = 0 \dots  L_y\! - \!2\}$. 
Therefore $\bar{D}=D_0 \cup D_{L_y}$, where $D_0 \equiv \{\sqrt{2}(2n+1,0) , n = 0 \dots  \floor*{\frac{L_x}{2}}-1  \}$; $D_{L_y} \equiv \{\sqrt{2}(2n+1,L_y-1) , n = 0 \dots  \floor*{\frac{L_x}{2}}-1  \}$ if $L_y$ is odd and $D_{L_y} \equiv \{\sqrt{2}(2n,L_y-1) , n = 0 \dots  \floor*{\frac{L_x}{2}}-1  \}$ if $L_y$ is even.
Now we see there are $O(L_x)$ sites in $\bar{D}$ that one can freely assign $|s_x = \pm1 \ra$. 
These states are apparently orthogonal and there are $O(2^{L_x})$ of them.
We therefore conclude that there are at least $O(2^{\max (L_x,L_y)})$ states of the type $|V(D),\{s_x\} \ra$ described in the main text.

\section{Exact Scar States of the PXP model on a Cubic lattice}

Based on the dimer construction on a square lattice, we can also construct exponentially many exact zero-energy states of the PXP model defined on a 3-dimensional cubic lattice even without periodic boundary condition. 
Consider a cubic lattice of size $N= L_x \times L_y \times L_z$ with $L_i$ even and open boundary conditions. 
The lattice sites are labeled by $\vec{r}=(r_x,r_y,r_z)$, where $r_i = 0,\dots, L_i-1$. 
The sites are partitioned again into sublattices $A = \{ \vec{r} \,\, | r_x+r_y +r_z \in \text{even}\}$ and $B = \{ \vec{r} \,\, | r_x+r_y +r_z \in \text{odd}\}$. 
For later convenience we also define $A_{E} = \{\vec{r} \,\, | r_{x,y,z} \in \text{even}\}$. 

Motivated by the eigenstates on a square lattice, we freeze the $B$ sublattice by putting a state on the $A$ sublattice that always have at least one $\ket{1}$ state adjacent to every site in $B$. 
We also need that the state on $A$ to be annihilated by the PXP Hamiltonian $H_{\text{PXP}}$. 

First, consider the $2\times 2\times 2 $ fundamental cube $C=\{\vec{r} \,\, | r_i \in \{0,1\}\}$. 
If two sites of the $A$ sublattice are in $\ket{1}$ then the sublattice $B$ is already frozen. 
Notice that in $C$, every point on the $A$ sublattice is adjacent to all points in the $B$ sublattice except for the diagonal point. This nonadjacent point is unique for each point in $A\cap C$. 
Therefore all the points on $B \cap C $ are adjacent to at least one point of every two points in $A \cap C$, which implies that any dimer covering on $A$ sublattice will freeze the $B$ sublattice to $|0\ra$.
We then only need to find dimer states that are annihilated by the PXP Hamiltonian. In the $B$-frozen subspace, $H_{\text{PXP}}$ is just $\sum_{\vec{r}\in A}X_{\vec{r}}$ so it will annihilate any total spin $S=0$ state (interpreting the two level systems as ${SU}(2)$ variables). 
Guided by the usual decomposition of $SU(2)$ representations:
\begin{equation}
    \frac{1}{2} \otimes \frac{1}{2} \otimes \frac{1}{2} \otimes \frac{1}{2} = 0 \oplus 0 \oplus 1 \oplus 1 \oplus 1 \oplus 2,
\end{equation}
there should be 2 different states that have total spin $S=0$. 
In fact, one of them can be readily seen as a dimer state from the fusion of the $SU(2)$ representation:
$\frac{1}{2} \otimes \frac{1}{2} = 0 \oplus 1$ and then one of the total spin $0$ representation is coming from $0 \otimes 0$. 
We write this state as 
\begin{align}
    | \boldsymbol{\chi}_{0}\rangle &= |\chi_{\vec{0},\vec{e}_x+\vec{e}_y}\rangle|\chi_{\vec{e}_x+\vec{e}_z,\vec{e}_y+\vec{e}_z}\rangle~,
\end{align}
where ${\vec{e}_i}$ are the unit vectors in the $i$-th direction.
While the other state can be found out by the similar procedure, we can in fact try to write down a different dimer state and see if they are linearly independent.
Indeed, consider
\begin{align}
| \boldsymbol{\chi}_{1}\rangle &= |\chi_{\vec{0},\vec{e}_y+\vec{e}_z}\rangle|\chi_{\vec{e}_y+\vec{e}_x,\vec{e}_z+\vec{e}_x}\rangle~,
\end{align}
and we have $\langle \boldsymbol{\chi}_{0} | \boldsymbol{\chi}_{1} \rangle = -\tfrac{1}{2}$.
The total spin $S=0$ space is therefore spanned by $| \boldsymbol{\chi}_{0}\rangle$ and $| \boldsymbol{\chi}_{1}\rangle$.

If we define $| \boldsymbol{\chi}_{\vec{r},\omega_{\vec{r}}} \rangle$ by displacing $|\boldsymbol{\chi}_{\omega_{\vec{r}}}\rangle$ by $\vec{r}$, the new zero energy scar states are given by 
\begin{equation}
    |{\Omega}_A;\{\omega_{\vec{a}}\}\rangle = \bigotimes_{\vec{a}\in A_E} | \boldsymbol{\chi}_{\vec{a},\omega_{\vec{a}}} \rangle \bigotimes_{\vec{b}\in B} | 0_{ \vec{b}}\rangle.
\end{equation}
We can similarly define similar states by switching the $A$ and $B$ sublattices. As there are 2 states for each fundamental cube, there are in total 
\begin{equation}
    N_{E=0,\text{Scar-3D},\text{OBC}}= 2 \cdot 2^{\frac{L_x L_y L_z}{8}}.
\end{equation} 
such states. Notice that even though we do not have the exact dimension of the PXP model on the cubic lattice, we have that 
\begin{equation}
     2^\frac{L_x L_y L_z+2}{2} -1\le \text{dim}\mathcal{H}_{PXP}^{(L_x,L_y,L_z)} \le 2^{L_x L_y L_z}
\end{equation}
for even $L_i$. The lower bound comes from the states with a given a sublattice froze to state $|0\rangle$ and the minus one is to not overcount the state with all $|0\rangle$'s. The upper bound comes from the total dimension of the unconstrained Hilbert space.

%\nocite{*}

\end{document}